\documentclass[11pt,a4paper]{article}

 \newcommand{\be}{\begin{equation}}
 \newcommand{\ee}{\end{equation}}
 \newcommand{\bl}{\begin{equation}\begin{array}{ll}}
 \newcommand{\el}{\end{array}\end{equation}}
 \newcommand{\bll}{\begin{equation}\begin{array}{lll}}
 \newcommand{\bdm}{\begin{displaymath}}
 \newcommand{\edm}{\end{displaymath}}
 \def\bea{\begin{eqnarray}}
 \def\eea{\end{eqnarray}}
 \def\barr{\begin{array}}
 \def\earr{\end{array}}
\def\p{\partial}
\def\d{\partial}

\def\f{\varphi}

\def\half{\frac{1}{2}}
\def\quart{\frac{1}{4}}

\def\third{\frac{1}{3}}
\def\2third{\frac{2}{3}}
\def\4third{\frac{4}{3}}
\def\3quart{\frac{3}{4}}

\def\sixth{\frac{1}{6}}

\def\lim{\rightarrow}

\def\bV{\bar{V}}
\def\bk{\bar{k}}

\def\bal{\bar{\alpha}}

\def\cA{{\cal A}}
\def\cF{{\cal F}}

\def\cL{{\cal L}}

\def\drho{\dot{\rho}}

\def\dsig{\dot{\sigma}}
\def\dpsi{\dot{\psi}}
\def\dal{\dot{\alpha}}
\def\dbe{\dot{\beta}}
\def\dga{\dot{\gamma}}
\def\dpsi{\dot{\psi}}
\def\dA{\dot{A}}
\def\dB{\dot{B}}
\def\dF{\dot{F}}
\def\df{\dot{f}}

\def\Lag{{\cal L}}
\def\Hag{{\cal H}}

\begin{document}
\raggedbottom

\title{{\bf On Einstein - Weyl unified model of \\ dark energy and dark matter}}

\author{A.T.~Filippov \thanks{Alexandre.Filippov@jinr.ru}~ \\
{\small \it {$^+$ Joint Institute for Nuclear Research, Dubna, Moscow
Region RU-141980} }}

\maketitle

\begin{abstract}

 Here we give a more detailed account of the part of
 the conference report\footnote{
`Selected problems of modern theoretical physics' Dubna, Russia,
 June~23-27, 2008} that was devoted to
 reinterpreting the Einstein `unified models of gravity and
 electromagnetism' (1923) as the unified theory of dark energy
 (cosmological constant) and dark matter (neutral massive
 vector particle having only gravitational interactions).
 After summarizing Einstein's work and related earlier
 work of Weyl and Eddington we present an approach to finding
 spherically symmetric solutions of the simplest variant of
 the Einstein models that was earlier mentioned in Weyl's work
 as an example of his generalization of general relativity.
 The spherically symmetric static solutions and homogeneous
 cosmological models are considered in some detail.
 As the theory is not integrable we study approximate solutions.
 In the static case, we show that there may exist two horizons and
 derive solutions near the horizons. In cosmology, we
 propose to study  the corresponding expansions of possible
 solutions near the origin and derive these expansions in a
 simplified model neglecting anisotropy.
 The structure of the solutions  seems to hint at a possibility of
 an inflation mechanism that does not require adding scalar fields.

\end{abstract}

\section{Introduction}
In this report I give a new interpretation of the `unified theory  of
 gravity and electromagnetism' proposed by A.Einstein in 1923
 in \cite{Einstein1} and briefly summarized in \cite{Einstein2}.
 Einstein gave no details of his derivations, presented no exact
 or approximate solutions, and did not explain why he completely
 abandoned his theory (I failed to find any reference to his papers
 \cite{Einstein1} - \cite{Einstein2} in his later work). Apparently
 these papers were soon forgotten by the scientific community and I could
 not find any reference to these papers in the second half of the 20-th
 century except for interesting remarks
 by Schr\"odinger \cite{Erwin} and a critical discussion by Pauli in
 addenda to the English translation of his famous book \cite{Pauli}.
 For these reasons, I first give a brief historical introduction
 summarizing Einstein's ideas and results as well as earlier
  related work of Weyl and Eddington.

 Immediately after the general relativity was formulated in its
 final form (1915 -1916) some attempts to modify it started.
 Einstein himself
 added the cosmological constant term $\Lambda$ to save (unsuccessfully)
 his static cosmology. After Friedmann's work (1922-1924) this
 modification was becoming more and more dubious. Weyl, after 1918,
 developed a much more serious modification aimed at unifying
  gravity and electromagnetism (most clearly summarized in \cite{Weyl},
  see also Einstein`s own summary \cite{Einstein3}).
  Starting from Levy-Civita's ideas
  on a general (non-Riemannian) connection (1917) he developed the theory of
  a special space in which the connection depends both on metric
 tensor and on a vector field which he tried to identify with
     the electro-magnetic
 potential. To get a consistent theory he introduced a general idea of
 gauge invariance which survived although the theory itself failed
 as he admitted later. In paper \cite{Einstein3} Einstein
 discussed Weyl's theory and
 expressed (like Pauli in \cite{Pauli}) the opinion that the theory
 is mathematically
 very interesting but probably not physical, at least, in its
 original formulation.

 In 1919 Eddington proposed a more radical modification of general
 relativity \cite{Ed1}, \cite{Ed}. His idea was to start with the pure affine
 formulation of the gravitation, i.e. using first the general symmetric
 affine  connection and only at some later stage
 introducing a metric tensor. Indeed, the curvature tensor can be defined
 without metric (here we use Einstein's notation \cite{Einstein1}
 but denote differentiations by commas):
 \be
 \label{1}
 r^i_{klm} = -\Gamma^i_{kl,m} + \Gamma^i_{nl} \Gamma^n_{km}
 + \Gamma^i_{km,l} - \Gamma^i_{nm} \Gamma^n_{kl} \,.
 \ee
 Then the Ricci-like (but non-symmetric) curvature tensor can
 be defined by contracting the indices $i, m$ (or, $i, l$):
 \be
 \label{2}
 r_{kl} = -\Gamma^m_{kl,m} + \Gamma^m_{nl} \Gamma^n_{km}
 + \Gamma^m_{km,l} - \Gamma^m_{nm} \Gamma^n_{kl}
 \ee
 (let us stress once more that $\Gamma^m_{nl} = \Gamma^m_{ln}$
 but $r_{kl} \neq r_{lk}$). Using only these tensors and the
 anti-symmetric tensor density one can build up a rather
 rich geometric structure. In particular, Eddington discussed
 different sorts of tensor densities \cite{Ed}.
 A notable scalar density is
 \be
 \label{3}
 \hat{\Lag} \equiv \sqrt{ -\textrm{det}(r_{kl})} \,
 \equiv \, \sqrt{ -r}
  \ee
 which resembles the fundamental scalar density of the Riemannian
 geometry, $\sqrt{-\textrm{det}(g_{kl})} \equiv \sqrt{-g}$.
    For this and some other
 reasons Eddington suggested to identify the
 symmetric part of $r_{kl}$ with the metric  tensor.
 The anti-symmetric part,
 \be
 \label{4}
 \phi_{kl} \equiv \half (\Gamma^m_{km,l}  - \Gamma^m_{lm,k}) \,,
  \qquad  \phi_{kl,m} + \phi_{lm,k} + \phi_{mk,l} \equiv 0\,,
 \ee
 strongly resembles the electro-magnetic field tensor and it seems
 natural to identify it with this tensor.
 Eddington tried to write consistent equations of
 the generalized theory
 but this problem was solved only by Einstein.

 The starting point of Einstein in his first paper
   of the series  \cite{Einstein1}
 was to write the action principle and to suppose (\ref{3}) to be
 the Lagrangian density depending on 40 connection functions
 $\Gamma^m_{kl}$.
 Varying the action w.r.t. these functions
 he derived 40 equations that allowed him to find the general
 expression for $\Gamma^m_{kl}$
 (the derivation is similar to that of
 the standard general relativity):
 \be
 \label{5}
 \Gamma^m_{kl} = \half [s^{mn} (s_{kn,l} + s_{ln,k} - s_{kl,n}) -
 s_{kl} \, i^n + {\third} ( \delta^m_k \, i_l +  \delta^m_l \, i_k)] \,.
 \ee
 Here $s_{kl}$ is a symmetric tensor ($s^{mn}$ is the inverse matrix
 to $s_{kl}$), which Einstein interpreted as
 the metric tensor (then the first term is the Christoffel symbol
 for this metric), and $i_n$ is a vector which he tried to
 connect with the electro-magnetic field. This
 identification apparently follows from the equations
 \be
 \label{6}
 r_{kl} = R_{kl} + \sixth [(i_{k,l} - i_{l,k}) + i_k i_l] \,,
 \ee
 \be
 \label{7}
 \phi_{kl} = \sixth (i_{k,l} - i_{l,k}) \,,
 \ee
 which can be obtained by inserting the expression (\ref{5})
 into (\ref{2}), (\ref{4}); $R_{kl}$ is the standard Ricci
 curvature tensor for the metric $s_{kl}\,$. Einstein's interpretation
 of $\phi_{kl}$ as the Maxwell field is not so natural because of the
 term $i_k i_l$ in the r.h.s. of Eq.(\ref{6}) which in fact makes this
 interpretation impossible. First, this term is not gauge invariant
 (but the gauge invariance was not yet discovered, the first clear
 formulation of the gauge principle was given by V.Fock in 1926).
 For Einstein, the main problem was that the electro-magnetic field
 in this theory could not exist without charges (i.e. there is no
 free field). To solve this problem he suggested to make this term
 `infinitesimally small' by choosing the corresponding dimensional
 constant (above, we omit all dimensional constants that can easily be
 restored). But we, today, cannot be satisfied with this solution
 because this term violates  gauge invariance and  makes the
 photon effectively massive (while it is known that
 there exist no continuous transition
 from the massless to massive photon theory)\footnote{
 The present best {\bf experimental} upper bound on the photon mass is
 $m_{\gamma} < 10^{-51}$g \cite{gamma}. Theoretical wisdom says that
 it must be zero.}.
   We return to discussing
  these facts, on which our interpretation of the Einstein theory
  is based, after considering the final proposals of Einstein.

  In his first paper (\textit{`Zur allgemeinen Relativit\"atstheorie'}),
  he considered two limiting cases. He showed that,
  when the $i_n$-terms in the connection vanish, the theory is
  equivalent to the standard general relativity with the cosmological
  term that emerges naturally and cannot be removed. In the flat space
  limit he demonstrated that weak fields $\phi_{kl}$
  (linear approximation) satisfy the free Maxwell equations
  provided that the $i_m i_n$-terms can be neglected.
  In the second paper of the series \cite{Einstein1} he gave the following
  expression for the effective Lagrangian density:
 \be
 \label{8}
 \hat{\Lag} = -2\sqrt{ -\textrm{det}(r_{mn})} + \hat{R}
 - \sixth \hat{s}^{mn}  i_m i_n \, .
  \ee
 This should be varied w.r.t. $\hat{s}_{kl}$ and
 $\hat{f}_{kl}$, which are {\bf the tensor densities
 defined with the aid of the
 scalar density} $\sqrt{-\textrm{det}(s_{kl})}$
 and corresponding to the tensors in the decomposition,
 \be
 \label{9}
 r_{kl} = s_{kl} + \phi_{kl} \,;
 \ee
 $\hat{R}$ is the scalar curvature density for the metric $s_{kl}$.
 The Lagrangian (\ref{8}) contains a very complex term
 $\sqrt{-\textrm{det} r_{mn}}$ which is more general than the
 so called Born-Infeld Lagrangian proposed ten years later
 \cite{Born} (the first attempts to construct nonlinear
 electro-dynamics were undertaken in \cite{Mie}).
 Apparently, Einstein did not try to find any particular solution of
 this theory and, instead, in the beautiful third paper
 `\textit{Zur affinen Feldtheorie}' (\cite{Einstein1})
 he proposed a significantly simpler effective Lagrangian
 that is the main subject of this paper.
 As he mentioned in the first two papers the actual
 form of the Lagrangian is unimportant for getting the connection
 (\ref{5}),  the only important thing is on which variables it
 depends.

 The main idea of the third paper is to take
 for the Lagrangian $\hat{\Lag}$ an arbitrary function of
 $s_{kl}$ and $\phi_{kl}\,$.\footnote{
 In his previous work Einstein implied that $\hat{\Lag}$
  depends on $r_{kl}\,$.
 At this point he quoted an unpublished work of
  `Droste (from Leiden)' who `two years ago expressed similar
  views'. The meaning of this rather cryptic remark is clarified
  in the second brief paper of ref.  \cite{Einstein2}, where he confirms that
  (Johannes) Droste  proposed to use a similar effective Lagrangian
  and, possibly a similar model (maybe,
  without cosmological constant).}
  Then he  introduces the Legendre transformation and
  the transformed (effective) Lagrangian
  density $\hat{\Lag}^*$:
 \be
 \label{10}
  {{\p \hat{\Lag}} \over {\p s_{kl}}} \equiv \hat{g}^{kl} \,, \qquad
 {{\p \hat{\Lag}} \over {\p \phi_{kl}}} \equiv \hat{f}^{kl} \,; \qquad
 s_{kl} = {{\p \hat{\Lag}^*} \over {\p \hat{g}^{kl}}} \,, \qquad
 \phi_{kl} = {{\p \hat{\Lag}^*} \over {\p \hat{f}^{kl}}}
  \ee
 Introducing the Riemann metric tensor $g_{kl}$ and the $i^k$-vector,
\be
 \label{11}
 g^{kl} \sqrt{-g} \, = \, \hat{g}^{kl} \,, \quad
 g_{kl} \, g^{lm} \, = \, \delta^m_k \,; \qquad
 \hat{i}^k = \p_l \hat{f}^{kl} \, ,
\ee
 he claims (without proof) that Eq.(\ref{5}) is valid with $s_{kl}$
 replaced by $g_{kl}$ and thus the affine geometry is
 the same for any $\hat{\Lag}(s_{kl},\phi_{kl}) $.
  Finally, he uses the freedom in choosing
 $\hat{\Lag}^*(\hat{g}^{kl}, \hat{f}^{kl})$ and proposes
  the following effective Lagrangian density:
 \be
 \label{12}
 \hat{\Lag}^* \, = \, 2\alpha \sqrt{-g} -
 \half \beta f_{kl} \, \hat{f}^{kl} \,,
 \ee
 where $\alpha$ and  $\beta$ are some constants not
 defined by the theory. This Lagrangian incorporates
 main properties of the theory discussed in two previous papers
 but is easier to deal with.

 To further clarify the relation of the new theory to
 general relativity Einstein rewrites the Lagrangian so that
 the equations of motion can be obtained by varying it in the metric
 and the vector field tensors, $g_{kl}$ and $f_{kl}\,$. Neglecting
 dimensions (for example, taking $\hbar = c = \kappa =1$) and
 changing Einstein's notation we write it as follows:
 \be
 \label{13}
 \hat{\Lag}\, = \,  \sqrt{-g}\, \bigl[R - 2\Lambda - \half F_{kl} F^{kl}
 - \mu^2 A_k A^k \bigl] , \qquad  F_{kl} \equiv A_{k,l} -  A_{k,l} \,.
 \ee
 Now it is absolutely evident that the vector field $A_k$ is not
 the Maxwell field.\footnote{
 Einstein tried to identify $A_k$ with a `cosmic current'
 (this explains his notation $i_k$).  A similar identification
 reappeared much later in quantum field theory
 (in the vector dominance model) under the name
 `field-current identity'. However, it is meaningless in the classical
 Maxwell theory.}
 Obviously, $A_k$ is a neutral massive vector field with coupling
 to gravity only. We will call it {\bf vecton}, that is an old fashioned
 but proper term for this `geometric' particle. This particle has not
 been directly observed but it can be considered as one of the possible
 candidates for {\bf dark matter}.  In view of the fact that
 the affine theory also predicted the cosmological constant term which
 is one of the best candidates for explaining {\bf dark energy},
 Einstein's theory may be considered as
  the first {\bf unified model of dark energy and dark matter}.

 Before we turn to further study of this model let us finish
 our presentation of its history. If you compare the Einstein
 model with the concrete models proposed in Weyl's book \cite{Weyl},
 you will find that Lagrangian similar to Eq.(\ref{13}) is one of Weyl's
 examples. Einstein's and Weyl motivations and approaches were quite
 different, and the Weyl connection does not coincide with Eq.(\ref{5})
 (see Addendum). Weyl's approach was mostly geometrical and
 he wrote the Lagrangian as a simplest illustration of possible
 physical applications, responding to criticism  by Einstein, Pauli and
 other physicists. Einstein was most interested in physics and, especially,
 in cosmology. Weyl criticized Einstein for his departure from geometric
 foundations of physics, in particular, for his derivation of geometry
 from the variational (action) principle which, probably, was his main
 achievement in the third paper. Note also that Weyl included the
 cosmological term only to avoid contradiction to Einstein cosmology
 of that time (`before Friedmann') while in the original Einstein model
 (\ref{3}), (\ref{7}) it was unavoidable. I think that, conceptually,
 the model (\ref{13}) is a step backwards,  in comparison with the original
 theory, (\ref{3}), (\ref{8}).
 There were, probably, two reasons for this step. First, Einstein's
 deep belief in simplicity of fundamental laws
 (`...aber \textit{boshaft ist Er nicht}').
 Second, his disappointment\footnote{
 In 1917 de~Sitter discovered a non-static solution of the empty - space
 Einstein equations with the cosmological constant.
 In 1923 Weyl and Eddington found the effect of recession of test particles
 in the de~Sitter universe. Thus $\Lambda$ was becoming useless
 and Einstein finally dismissed it in 1931, after Hubble's discovery (1929).}
 in static cosmology after
 accepting Friedmann's results, \cite{Fried}. Anyway, in his
 last papers on affine theory \cite{Einstein2} he set the cosmological
 term to zero what is impossible in the original theory and
 quite unnatural in the framework of the affine approach.

 Above, we also mentioned
 work and ideas of Eddington. The intensive exchange of ideas
  between Einstein, Weyl and Eddington resulted in interrelations
  in their work (published in 1918-1923) that are difficult
  (and, possibly, unnecessary) to disentangle. As the constructive
  ideas of the affine theory were mostly created by Weyl, Eddington,
  and Einstein, the resulting model should probably be called
  {\bf Einstein-Weyl-Eddington unified model of dark energy and
  dark matter}. However, as far as I am here discussing the concrete
  Lagrangian (\ref{13}), I call it Einstein-Weyl model.

  Before turning to new results let us briefly summarize the results
  and thoughts of Weyl, Eddington, and Einstein.
  {\bf 1.~Weyl} had a very clear and original geometric ideas, but: a) his
physics was rightly criticized by Einstein, Pauli, and other physicists,
b) he considered the theory as a unified theory of gravity and
electromagnetism but his vector field was also not electro-magnetic, c)
his discussion of dynamics was incomplete and he himself regarded it as
preliminary. Nevertheless, it is possible that not all the potential of
the Weyl ideas is understood and used.
  {\bf 2.~Eddington} proposed to
use, instead of the Weyl's non-Riemannian `metrical spaces',  the most
general spaces with symmetric affine connection (without torsion). He
discussed possible invariants that can be used in physics, in particular,
the square root of the determinant of $R_{kl}\,$.\footnote{
 In addendum to \cite{Ed}, where he gave a clear and detailed account of
 the Einstein work \cite{Einstein1}, he also discussed another scalar
 density that was proposed by R.~Weitzenb\"ock}
 He proposed to consider the symmetric part of the curvature matrix
as the metric in the general space and the anti-symmetric one as the
electro-magnetic field tensor. In later works he discussed a possibility
to use this as a Lagrangian (long before the proposal of Born and Infeld).
However, he did not find a consistent approach to dynamics.
 {\bf 3.~Einstein} started with formulating dynamics by use of the Hamilton
principle similar to one proposed by Palatini in general relativity. The
new (and crucial) idea was not to introduce any metric at the beginning
and not to fix any special form of the affine connection (apart of the
symmetry condition). He soon realized
 (in the second apper), that he does not need to use a concrete form of
 the Lagrangian that
can be just any function (tensor density) of
 $s_{kl}$  and $\phi_{kl}$-matrices (see (\ref{9}-\ref{11})).
 For any such Lagrangian he proved that the affine
 connection allows one to introduce a symmetric metric and found
 the expression
 for connection. Both Einstein's and Weyl's expressions are special cases
 of the general formula for the symmetric connection (see Appendix).

 The most important thing is the following:
 supposing that the equations of motion follow from an action principle
 with the general Lagrangian fixes the geometry (connection) and,
 eventually, allows one to fix some metric compatible with this
 non-Riemannian connection. Another important thing is that
 the action can be
 (and should be) written without metric. Using the
 Legendre transformation Einstein bypassed difficulties that were met
 on this way and wrote more tractable effective Lagrangian, but some
 conceptually beautiful and
 important features of his new theory were thus hidden (or even lost).

 Apparently, Einstein was disappointed in the cosmological constant and
 also gradually realized that his interpretation of the anti-symmetric
 field as the electro-magnetic field was not quite satisfactory. Anyway, he
 completely abandoned this model and left no detailed
 account of his work.
 He did not mention any static or cosmological solutions
 even in the simplified version of the theory, (\ref{13}). In this paper
 we try to fill this gap and establish grounds for comparing this model
 to the present day cosmology.

 \section{Spherical reduction - static and cosmological solutions}
 \subsection{Vecton-dilaton gravity}
 At first sight, the theory (\ref{13}) is very close to the
 well-understood Einstein-Maxwell theory which can be obtained
 when $\mu = 0$. However, we will show that the two theories are
 qualitatively different and it is hardly possible to construct
 a reasonable perturbation theory in the parameter $\mu^2$.
 We start our qualitative analysis without assuming that this
 parameter is small. The natural object for this analysis is
 the spherically reduced theory. When $\mu = 0$, the theory automatically
 reduces to rather simple one-dimensional equations that can be
 explicitly solved. The solution is the Reissner - Nordstr\"om
 black hole (when the electric charge vanishes it reduces to the
 Schwarzschild black hole). In general, when gravity couples to
 other (not electro-magnetic) fields the spherically reduced theory
 is described by two-dimensional differential equations which
 are not integrable except very special cases
 (for many examples and references see, e.g.,
 \cite{Kummer}-\cite{ATF5}.).

 Following the approach to dimensional reduction and to
 resulting 1+1 dimensional dilaton gravity (DG)
  developed in papers \cite{ATF1}-\cite{ATF4}
 it is not difficult to derive these equations.
 The general spherically symmetric metric is
 ($i,j = 0,1$; $x^0 = t, x^1 = r$):
 \be
 \label{14}
 ds^2 = g_{ij}(t,r)\ dx^i dx^j + \f(t,r) ( \sin^2 \theta \, d\theta
 + d\phi^2)\,.
 \ee
 Supposing that all other functions also depend on $t,r$, inserting the
 metric (\ref{14}) into the action with the Lagrangian (\ref{13}),
 and integrating out the angle variables $\theta, \phi$ one can
 derive the following effective two-dimensional Lagrangian\footnote{
 Very similar effective Lagrangians can be obtained from the
 higher-dimensional analogs of the Lagrangian (\ref{13}).
 On the (1+1)-dimensional level it is not difficult to include other
 sorts of matter that appear in reductions of higher-dimensional
 supergravity theories (for references see, e.g., \cite{ATF1} -
 \cite{ATF5}). I hope to discuss some of these generalizations in
 future publications.}:
\be
\label{15}
 \Lag^{(2)} \, = \,  \sqrt{-g} \, \bigl[\f \, R^{(2)} + 2 - 2\Lambda \f +
 (\p \f)^2/2 \f
 - \f \, F_{ij} F^{ij} - \f \, \mu^2 A_i A^i \bigl] ,
\ee
 where $R^{(2)}$ is
 the two-dimensional Ricci curvature depending on the $g_{ij}\,$ (the second term in the brackets is the 3-space curvature).
 It is convenient to remove the fourth term by the Weyl rescaling
 of the metric, $g_{ij}= \f^{-\half} g^{\textrm{w}}_{ij}\,$.
 Below we use the transformed Lagrangian,
 \be
\label{16}
 \Lag_W^{(2)} \, = \,  \sqrt{-g} \, \bigl[\f \, R^{(2)} + 2\f^{-1/2} -
 2\Lambda \f^{1/2} - \f^{-3/2} \, F^2 - \f \, \mu^2 A^2 \bigl] \,.
\ee

 It is easy to derive the equations of motion which in
 a generic metric $g_{ij}\,$ are equivalent to the Einstein equations
 for the spherically symmetric solutions of the four-dimensional theory
 (\ref{13}). By varying w.r.t. the diagonal metric functions $g_{ii}\,$
 we fist write the energy and momentum constraints. In the light cone (LC)
 metric, $ds^2 = -4 f(u,v)\, du \,dv$, these constraints are simple:
 \be
 \label{17}
 f \p_i \, (\p_i \,\f /f) + \f \, \mu^2 A^2_i = 0, \qquad
 \qquad  i = u,v \,.
 \ee
 The constraints (\ref{17}) should be derived using the general
 metric $g_{ij}\,$.
 The other equations of motion may be obtained directly in the LC-metric:
  \be
 \label{18}
 \p_u \p_v \, \f \, + \, f \bigl(2\f^{-1/2} -  2\Lambda \f^{1/2} -
 \half \f^{3/2} \, f^{-2} F^2_{uv} \bigr)=0, \qquad F_{uv} \equiv
 A_{u,v} - A_{v,u} \, ,
 \ee
 \be
 \label{19}
 \p_j \, \bigl(\f^{3/2} f^{-1} F_{ij}\bigr) =  \f \, \mu^2 A_j \,
 \qquad   i,j = u,v \, .
 \ee
 From the last equation immediately follows that
 $\p_v (\f A_u) + \p_u (\f A_v) = 0$ .
 In the original four-dimensional theory this is the
 $\p_{\mu} (\sqrt{-g} A^{\mu}) = 0$ condition eliminating
 spin 0. Weyl, Eddington and Einstein called it the
 Lorentz condition although we know that its origin and
 meaning are quite different from the gauge fixing condition
 in the Maxwell theory first introduced by L.Lorenz and
 later popularized by H.A.Lorentz.

 This dilaton gravity coupled to massive vector field
 (I suggest to call it {\bf vecton-dilaton gravity, VDG})
 is more complex than the well
 studied models of dilaton gravity coupled to scalar fields and
 thus it requires a separate study.
 The natural first question is: are there exact
 analytical solutions like Schwarzschild or Reissner-Nordstr\"om black
holes? If the vector field is constant, we return to exactly soluble DG
having explicit solutions with horizons. Otherwise, when the vector field
is nontrivial, the answer is more difficult to find but it is worked out
in some detail below.
 The second question is: what are the simplest cosmological solutions in
 this theory?
 Thus, the  first thing to do is to further reduce the theory to static
or cosmological configurations. Consider first the static reduction.

\subsection{Static states and horizons}
  The simplest way to derive the corresponding equations is to suppose
that all the functions in the equations depend on  $r = u+v$ . But this
  is not the most general dimensional reduction of
  the two-dimensional theory.
 There exist more general ones that allow us to simultaneously treat
 black holes, cosmologies and some waves.
 These generalized reductions were proposed in papers \cite{ATF6},
 \cite{ATF3}, \cite{ATF5}
devoted to dilaton gravity coupled to scalar fields and Abelian gauge
fields; here we only discuss in some detail the static and cosmological
reductions.  In both cases it can be seen that the perturbed theory
 (with a nonvanishing mass term) is qualitatively different from
 the non-perturbed
 one. Indeed, the non-perturbed theory is just dilaton gravity coupled to
  electromagnetism. This model is equivalent to pure dilaton gravity, which
is a topological theory. In particular, it automatically reduces to
one-dimensional static or cosmological models that can be analytically
solved. Static states are the Reissner-Nordstr\"om black holes perturbed
by the cosmological constant and having two horizons, while the space
between horizons may be
 considered as an unrealistic cosmology.
 This object is known from 1916 times;
 certainly it was familiar to Einstein in 1923 but he did not discuss
the static configuration and apparently did not consider black holes or
horizons as having any relation to physics.

 Let us now write the {\bf static} equations corresponding to the
 naive reduction
 to one spatial dimension. To obtain them one can reduce either the
 equations or the Lagrangian.
 Following \cite{ATF4}, \cite{atfm},
 we write the equations of motion  in a somewhat unusual form.
 Let us define two additional
 functions, $\chi$ and $B$, by the equations (the prime denotes
 differentiations w.r.t. $r$)
 \be
 \label{20}
 \f'(r) = \chi (r),  \qquad A'(r) = f(r) \f^{-3/2}(r) B(r)  \, ,
 \ee
 where, as follows from Eq.(\ref{19}), $A_v(r) = -A_u(r) \equiv -                                                                                                                                   A(r)$.
 Then the other equations are
 \be
 \label{21}
  \chi' = -fU \,,  \qquad B' = -\half \f \mu^2 A \,,  \qquad
  f' = (f/\chi) [-f U + \f \mu^2 A^2]  \, ,
 \ee
 where we defined the potential \be
 \label{22}
 U \equiv 2(\f^{-1/2} - \Lambda \f^{1/2} -
 \f^{-3/2} B^2) \, ,
 \ee
 {\bf These equations are not integrable and cannot be solved
 analytically.}
 To get numerical solutions we first have to study the analytic and
 asymptotic properties of their solutions.

 Here we only consider
 {\bf solutions near possible horizons} that are defined as zeroes of the
 metric, $f\rightarrow 0$ for finite values of
 $\f \rightarrow \f_0 \, $. It is not difficult to understand that
 we also should require that $A$ is finite near the horizon.
 To study the behaviour of the solutions for small values of
 $\tilde{\f} \equiv \f - \f_0 \,$
 it is most convenient to
  {\bf consider the solutions as functions of}
  $\f$. Further analysis shows that the solutions can
  be expanded in power series of  $\tilde{\f}$ and that
 the functions  $\tilde F \equiv f/\chi$ and $\tilde A = A/\chi$
 should be finite.
 Thus we have:
 \be
 \label{23}
 \tilde{F}'(\f) = \f \, \tilde{F}(\f) \, \mu^2 \tilde{A}^2(\f)  \,,
 \qquad  \qquad \quad
 \chi'(\f) = - \tilde{F}(\f) \, U(\f) \,, \qquad
 \ee
\be
 \label{24}
 B'(\f) = -\half \f \, \mu^2 \tilde{A}(\f) \,,  \qquad
 \tilde{A}'(\f) \chi (\f) = \tilde{F}(\f) \,
 \bigl[\f^{-3/2} B(\f) + U(\f) \tilde{A}(\f)\bigl] \,,
 \ee
 where now the prime denotes differentiation in the new variable $\f$.
 It is not very difficult to show that $\f_0$, $\tilde{A}_0$, $B_0$,
 $\tilde{F}_0$ can be taken arbitrary up to one relation that should
 be satisfied due to the second equation (\ref{24}):
 \be
 \label{25}
\tilde{A}_0 \, U_0 + \f_0^{-3/2} B_0 = 0 \,, \qquad
 \qquad U_0 \equiv U(\f_0, B_0) .
 \ee
This equation can be solved w.r.t any parameter. It is interesting
 to see that it has two solutions for $\f_0$ which means that
 {\bf there may exist two horizons}\footnote{
 This is similar to the charged Reissner-Nordstr\"om black
 hole or to black holes in higher - dimensional theories, \cite{Stelle},
 although in the present model there is no conserved electric charge.}
 as distinct from the Schwarzschild
 black hole. Note that the solutions with different $\tilde{F}_0$
 are equivalent  because  the equations are invariant under the
 scale transformation $\tilde{F} \Rightarrow C\tilde{F}$,
 $\chi \Rightarrow C \chi$.

 Now, following the method of \cite{atfm}, one can find several
 terms in the expansion of the solution.
 Unfortunately, it is not clear how to construct the complete
 expansion and therefore our derivations do not allow us  to study
 global properties of the solutions.
 They say nothing about asymptotic properties
 and singularities which should be the subject of separate
 investigations.\footnote{
 The asymptotic expansions for $r \rightarrow 0$ and
 $r \rightarrow \infty$ can be obtained by following the
 approach proposed below for the cosmological solutions.}
 When the qualitative properties
 of the black hole type solutions will be understood, the
 static solutions and their formation can be studied
 by numerical simulations. As far as I know, the coupling of massive
 neutral vector particles to gravity did not attract much attention
 (see, however, numerical simulations of the critical collapse
 of a massive vector field in \cite{Mann}).

 \subsection{Cosmology}
 \subsubsection{General formulation}
 Let us turn to cosmological reductions. The simplest cosmology
 can be obtained by the same naive reductions as was used for
 static states. However, this is not the most general dimensional
 reduction giving all possible spherically symmetric cosmological
 solutions (similarly, the above naive reduction does not give
 all possible static spherically symmetric solutions).
 A more general procedure is described in \cite{ATF3}.
 Following this procedure we return to the two-dimensional Lagrangian
 (\ref{15}) but add to it a scalar field $\psi$ that should represent
 `matter' in low dimensions. To simplify comparison to the standard
 cosmological solutions let us now use the $(t,r)$-coordinates
 and write the general spherically symmetric metric as:
 \be
 ds_4^2 = e^{2\alpha} dr^2 + e^{2\beta} d\Omega^2 (\theta , \phi) -
 e^{2\gamma} dt^2 + 2e^{2\delta} dr dt \, ,
 \label{eq1}
 \ee
 where  $\alpha, \beta, \gamma, \delta$ depend on $t$, $r$ and
 $d\Omega^2 (\theta , \phi)$ is the metric on the 2-dimensional sphere
 $S^{(2)}$. Then  the two-dimensional reduction of the four-dimensional
 EW theory coupled to a scalar field $\psi$ can easily be found
 (here the prime denotes differentiations in $r$ and the dot -
 in  $t$:
 \be
 \cL^{(2)} = e^{2\beta} \bigl[e^{-\alpha - \gamma} (\dA_1 - A_0')^2 -
 e^{-\alpha + \gamma} (\psi'^2 + \mu^2 A_1^2) +
 e^{\alpha - \gamma} (\dpsi^2 + \mu^2 A_0^2) -
 e^{\alpha +\gamma} (V + 2\Lambda)  \bigr] + \cL_{gr} \,,
  \label{eq2}
 \ee
 where $\psi = \psi(t,r)$, $V=V(\psi)$, $A_i = A_i(t,r)$,
 $\dA_1 - A_0' \equiv F_{10}$ and
 \be
 \cL_{gr} \equiv
 e^{-\alpha + 2\beta +\gamma} (2\beta'^2 + 4\beta' \gamma') -
 e^{\alpha + 2\beta - \gamma} (2\dbe^2 + 4\dbe \dal) +
  2k e^{\alpha + \gamma}
 \label{eq3}
 \ee
 is the gravitational Lagrangian, up to the omitted total derivatives
 that do not affect the equations of motion.
 Variations of this Lagrangian give all the equations of motion except
 one constraint,
 \be
 -{\dot{\beta}}^{\prime} - \dot{\beta} {\beta}^{\prime} +
  \dot{\alpha} {\beta}^{\prime} + \dot{\beta} {\gamma}^{\prime} \,\,
  = \,\, \half [\dot{\psi} {\psi}^{\prime} + A_0 A_1] ,
 \label{eq4}
 \ee
 which should be derived before we omit the $\delta$-term in the metric
 (taking the limit $\delta \rightarrow -\infty$). All other equations
 of motion can be obtained from the effective Lagrangian
 (\ref{eq2})\footnote{ The equations are equivalent to
 Eqs.(\ref{17}) - (\ref{19}) but are written in the Einstein frame
 and in the $(t,r)$ coordinates.}.

 Now, the distinction between {\bf static} and {\bf cosmological}
 solutions is in the dependence of their `matter' fields $A_i$ and $\psi$
 on the space-time coordinates.
 We call {\bf static} the solution for which $A_i=A_i(r)$ and $\psi=\psi(r)$.
 If $A=A_i(t)$ and $\psi=\psi(t)$ we call the solution {\bf cosmological}.
 There may exist also the {\bf wave-like} solutions for which
 $A$ and $\psi$ depend on linear combinations of $t$ and $r$ but here
 we do not discuss this possibility (see, e.g., \cite{ATF7} and references
 therein). For both static and cosmological solutions the gravitational
 variables in general depend on $t$ and $r$.
 To solve the equations of
 motion we may reduce them by separating $t$ and $r$.
 It is clear that to separate the variables $r$ and $t$ in the metric we
 should require that
 \be
 \alpha = \alpha_0(t) + \alpha_1(r) , \quad \beta = \beta_0(t) +
 \beta_1(r) , \quad \gamma = \gamma_0(t) + \gamma_1(r) , \,
 \label{eq5}
 \ee
 Inserting this into the equations of motion one can find the restrictions
 on the gravitational (and, possibly on the matter) variables that
 must be fulfilled. The details can be found in \cite{ATF3}, where one can
 find the complete list of the static and cosmological
 spherically symmetric solutions when
 the vector field identically vanishes (we call this case the `scalar cosmology).
 Here we only give a very brief summary
 and a simple generalization to nonvanishing vector field.

 The naive cosmological reduction (that supposes all the fields to be
 independent of $r$) does not give the standard FLRW scalar cosmology.
 As was shown in \cite{ATF6} and \cite{ATF3}
 (see also the earlier paper \cite{ATF1}),  all homogeneous isotropic
 cosmologies should satisfy the following conditions:
 \be
 \label{eq6}
 \dot{\alpha} = \dot{\beta} \,, \quad \gamma' = 0 \,, \quad
 {\beta_1}'' + k e^{-2 \beta_1} = 0 \,, \quad
     k e^{-2 \beta_1} \,-\, 3{\beta_1'}^2 - 2{\beta_1}'' \,=\, C \,,
 \ee
 where $C$ is a constant proportional to the 3-curvature
 (its time dependence is given by the factor $e^{-2 \alpha_0}$)
 and the third equation is the isotropy condition.
 Neglecting inessential constant factors, we also have chosen
 $\alpha_1 = \gamma_1 = 0$. We see that for the naive reduction
 the isotropy conditions in (\ref{eq6})
 can be satisfied only if $k = 0$ and
 that the first condition is not dictated by (\ref{eq4}).
 Thus, the naive reduction gives, in general, a homogeneous
 non-isotropic cosmology.

 For the FLRW {\bf scalar cosmology} $\beta' \neq 0$ and all the
 conditions (\ref{eq6}) must be satisfied. Then the effective
 one-dimensional Lagrangian that can describe both the naive and FLRW cosmology reads
 \be
 \cL^{(1)} = 6\bar{k} e^{\alpha + \gamma} -
 e^{2\beta} \bigl[e^{\alpha +\gamma} (V + 2\Lambda)  -
 e^{\alpha - \gamma} (2\dbe^2 + 4\dbe \dal - \dpsi^2) \bigr] \,,
 \label{eq7}
 \ee
 where $\bar{k}$ is a real constant related to $k$ and $C$;
 $\alpha, \beta, \gamma$, $\psi$ depend only on $t$, and $\gamma$
 is the Lagrangian multiplier (in cosmology, the standard gauge fixing is $\gamma = 0$).

 Taking now $\alpha(t) = \beta(t)$ we get
 the Lagrangian of the FLRW scalar cosmology, for which
 it is not difficult to derive the equations of motion.
 As in the static case, we can use $\alpha$ as the new
 independent variable and to derive the following first
 order equation for
 $\chi(\alpha) \equiv d\psi /d\alpha \equiv \psi'({\alpha})$:
 \be
 \frac{d\chi^2}{d\alpha} = (\chi^2 -6) \biggl[\chi^2 + \chi \frac{V_{\psi}}{V+2\Lambda}\biggr] \equiv (\chi^2 -6)\biggl[\chi^2 +  \frac{d}{d\alpha}\ln(V+2\Lambda)\biggr]\,,
 \label{eq8}
 \ee
 where we used the obvious relation $\psi'({\alpha})\,d/d\psi \equiv d/d\alpha$. This equation\footnote{It can be derived from Eq.(\ref{eq17}) and constraint (\ref{eq13}) below.} is valid if $\bar{k} = 0$, and it is independent of the gauge choice.  If we could analytically solve this equation, we would derive the expression for $\chi(\alpha)$ and thus for finding $\psi(\alpha)$ we may simply integrate $\chi(\alpha)=\psi'(\alpha)$ over $\alpha$. Then, using the constraint derived by differentiating the Lagrangian (\ref{eq7}) in the Lagrange multiplier $e^\gamma$ we find the Hubble function $\alpha(t)$ from equation (with $\gamma=0)$,
  \be
 \dal^2(t) = \bigl[V(\psi(\alpha)) + 2\Lambda\bigr] \,
 \bigl[6 - \chi^2(\alpha)\bigr]^{-1} \,,
 \label{eq9}\mathrm{}
 \ee
 which is the phase portrait, $\dal(\alpha)$ of the gravitational part of the cosmology.
 All this looks nice but does not give the desired analytical
 expressions although the above formulae may be useful for
 a qualitative analysis of possible solutions.  To get approximate analytical expressions it is better to rewrite (\ref{eq8}) as the equation for $\bar{\chi}(\psi)\equiv 1/\chi[\alpha(\psi)]\equiv d\alpha/d\psi$. Then we can derive asymptotic and power-series expansions of $\bar{\chi}(\psi)$ and thus, by transforming (\ref{eq9}) into the expression for the phase portrait of the scalar field $\dpsi(\psi)$, it is possible to find its behavior for large and small $\psi$.

 Returning to the vector model we first write the general cosmological EW Lagrangian supplemented
 by the minimally coupled scalar field (that may represent either
 matter or inflaton). At first sight, the dimensional reduction of
 the spherically symmetric Lagrangian (\ref{eq2})-(\ref{eq3})
 with the vector field must not differ from the usual one used for
 the scalar cosmology and can be written as:
 \be
 \cL^{(1)} = 6\bar{k} e^{\alpha + \gamma} +
 e^{2\beta} \bigl[e^{-\alpha - \gamma} \dA_1^2 -
 e^{-\alpha + \gamma} \mu^2 A_1^2 -
 e^{\alpha +\gamma} (V + 2\Lambda)  -
 e^{\alpha - \gamma} (2\dbe^2 + 4\dbe \dal - \dpsi^2) \bigr] \,,
 \label{eq10}
 \ee
 where all the fields depend on $t$.
 Then, taking $\alpha = \beta$, we apparently obtain
 a FLRW type cosmology with the vector field.
 However, unlike the scalar field,
 the two-dimensional vecton field equations,
 \be
 \label{ac4}
 \d_{\,0} [e^{\alpha_0 -\gamma_0 + 2\beta_1} \dA_1 ]
 = -\mu^2 e^{\alpha_0 + \gamma_0 + 2\beta_1} A_1 \,,
 \ee
 \be
 \label{ac5  }
 \ \d_{\,1} [e^{\alpha_0 -\gamma_0 + 2\beta_1} \dA_1 ]
 = -\mu^2 e^{3\alpha_0 - \gamma_0 + 2\beta_1} A_0 \,.
 \ee
  {\bf do give additional constraints} on $\beta_1(r)$.
 The first equation does not depend on $\beta_1$, but
 the second one requires either $\beta_1'(r) = 0$
 or $\beta_1'(r) = \textrm{const}$. The second condition
 gives $A_0 \sim \dA_1$ and so (\ref{eq4}) is incompatible
 with the isotropy condition $\dot{\alpha} = \dot{\beta}$.
  Therefore, there remains only the first case,
 $\beta_1'(r) \equiv 0$, from which it follows that
 $k  = \bk = 0$. Although the constraint (\ref{eq4}) is
 identically satisfied (as we suppose that $\gamma' = 0$)
 it does not give the necessary isotropy condition
 $\dot{\alpha} = \dot{\beta}$
 that automatically emerges in the scalar cosmology case.
 As we'll see in a moment, this condition cannot be
 exactly satisfied in the vecton cosmology and can only be
 approximate\footnote{This fact was not emphasized in the
 first version of this report. Moreover, the solutions
 derived there are only approximate  solutions of the
 four-dimensional theory as clearly follows from the
 general equations written below. The author is thankful to
 V.~Rubakov for calling his attention to this apparent inconsistency
 in the presentation of the results in the first version.
 }.

 Summarizing this discussion, we consider the vecton plus
 scalar cosmology described by the Lagrangian (\ref{eq10})
 with $k=\bk =0$ and $A_1$ being the $A_z$ component of the
 four-dimensional vector field (it follows that the
 cosmology must be in general non-isotropic).
 To write the corresponding equations of motion in a
 most clear and compact form we introduce the temporal
 notation
 \be
 \rho \equiv \third (\alpha + 2\beta) \,, \quad
 \sigma \equiv \third (\beta - \alpha) \,, \quad
 A_{\pm} = e^{-2\rho + 4\sigma} (\dA^2 \pm
 \mu^2 e^{2\gamma} A^2) \,, \quad
 \bV \equiv V(\psi) + 2\Lambda \,.
 \label{eq11}
 \ee
 where $A_1 \equiv A_z \equiv A$. Then the exact
 Lagrangian for EW plus scalar cosmology is:
 \be
 \cL^{(1)} = e^{3\rho - \gamma} (\dpsi^2 - 6\drho^2 + 6\dsig^2) +
 e^{3\rho - \gamma} A_{-}  -
 e^{3\rho + \gamma} \, \bV(\psi) \,.
 \label{eq12}
 \ee
 We see that $A, \psi, \rho, \sigma$ are the dynamical variables
 and $e^\gamma$ is the Lagrangian multiplier, variations of which
 give us the remaining energy constraint:
 \be
 \dpsi^2 - 6\drho^2 + 6\dsig^2 + A_{+} + e^{2 \gamma} \, \bV = 0
 \label{eq13}
 \ee
 (the momentum constraint
 (\ref{eq4}) is satisfied by construction). The other equations are:
 \be
 \ddot{A} + (\drho + 4\dsig - \dga) \dA +
 e^{2 \gamma} \mu^2 A = 0 \,,
 \label{eq14}
 \ee
  \be
 4\ddot{\rho} + 6\drho^2 -4\drho \dga  - 6\dsig^2 + \third A_{-}
 + \dpsi^2 - e^{2 \gamma} \, \bV = o\,,
 \label{eq15}
 \ee
 \be
 \ddot{\sigma} + 3 \dsig \drho - \dsig \dga -\third A_{-} = 0 \,.
 \label{eq16}
 \ee
 \be
 \ddot{\psi} + (3\drho - \dga)\dpsi +
 \half e^{2 \gamma} \, \bV_{\psi}  =  0 \,,
 \label{eq17}
 \ee
  These equations are much more complex than the equations
 of the scalar cosmology that can be obtained when $A \equiv 0$.
 The FLRW cosmology is obtained if in addition we choose $\sigma \equiv 0$.
 Even for $A \equiv 0$, there exist anisotropic solutions with $\sigma \neq 0$,
 but if the curvature parameter vanishes ($k=\bk =0$), they are probably unstable.  Cosmologists usually choose the gauge
   $\gamma = 0$. Here we also use the the light-cone (LC) gauge. A very useful gauge is $\gamma = 3\rho$, which simplifies the Hamiltonian; it is most useful in search of integrable theories.

 The above equations are {\bf not integrable} in any sense and rather difficult for a qualitative analysis.  Nevertheless, they are not much more difficult than the equations for the static solutions considered in Section~2.2 and we may apply the same approach to solving them in asymptotic regions. If the
 scalar field identically vanishes these equations are essentially
 identical to the equations for the static states.
  They also would be greatly simplified if it were possible to
 neglect the $\sigma$-field. Unfortunately, it is evident that
 this is in general impossible because then $A_{-} =0$ but this
 condition is incompatible with the other equations.
 This means that the exact solutions of the EW model (even with
 many scalar fields minimally coupled to gravity)
 should be {\bf non-isotropic}\footnote{
 If one would introduce other scalar fields non-minimally coupled to
 gravity, this statement may become not valid. At this stage of
 investigation, we are not ready to add other vector fields or
 fields with the the spin $1/2$.}.
 In the next subsection we consider the simplified model obtained
 by neglecting the anisotropy.

\subsubsection{Simplified model}
 To get the simplified model we neglect Eq.(\ref{eq17}) and
 take $\sigma \equiv 0$, $\psi \equiv 0$, $V(\psi) \equiv 0$.
 Then  $\rho = \alpha = \beta$ and the approximate effective Lagrangian
 (\ref{eq12}) becomes\footnote{
 Above we completely neglected the dimensions of all the variables
 and omitted the gravitational constant.
 Here we only need to restore one of the dimensions supposing that
 $[t^{-2}] = [k] = [\Lambda] = [\mu^2] = [L^{-2}]$. }
 \be
 \label{26}
 \Lag_a = - 6 \dal^2 e^{3\alpha - \gamma} -
 2\Lambda e^{3\alpha + \gamma} + \dA^2 e^{\alpha - \gamma}
 - \mu^2 A^2 e^{\alpha+\gamma} \,,
 \ee
 The corresponding equations of motion are the three equations
 (\ref{eq13})-(\ref{eq15}) with $\sigma = \psi = V =0$ and
 $\rho = \alpha$. The first equation, (\ref{eq13}), is
 equivalent to vanishing of the Hamiltonian.
 Denoting $f \equiv e^{\alpha}$ and taking
 {\bf the gauge fixing condition} $\gamma = 0$
 ({\bf the `standard' gauge})
 we have\footnote{
 Here we treat cosmological solutions independently of the
 static states and use somewhat different notation.
 For example, the normalization of $A$ and of $\mu^2$
 are slightly different. }
 \be
 \label{27}
 \Hag^a_0 \equiv f \bigl[-6 \df^2 +2\Lambda f^2 +
  \dA^2 + \mu^2 A^2 \bigr] = 0 \,.
 \ee
 Another useful {\bf gauge (the LC gauge) is}
 $\alpha = \gamma$.
 In this gauge, the effective Hamiltonian is:
 \be
 \label{28}
 \Hag^a_1 \equiv -6 \df^2 + 2\Lambda f^4 +
 \dA^2 + f^2 \mu^2 A^2 = 0 \,.
 \ee
 It is not difficult to write the equations independent
 of the gauge choice and we leave this as a simple exercise
 to the reader.

 Let us write the {\bf equations of motion in the LC gauge}
 $\alpha = \gamma$.
 In analogy to the static case we write them in the first order form
 (the first equation is the definition of $F$),
\be
 \label{29}
 F \equiv \dal = \df/f \,, \qquad \dF +F^2 = \2third \Lambda f^2 +\sixth \mu^2 A^2,
  \qquad \dA = B , \qquad \dB = -\mu^2 f^2 A \,,
 \ee
 and the Hamiltonian constraint is a simple polynomial function of
 $f, F, A, B$:
\be
 \label{30}
 \Hag^a_1 = -6 f^2 F^2 + B^2 +
 2\Lambda f^4 + \mu^2 f^2 A^2  =0 \,.
 \ee
 Similarly to our previous consideration of the static equations,
 we prefer to change the independent variable to
 $\alpha \equiv \ln f$.  It is convenient to introduce
 two new functions, $\psi (\alpha)$ and $G(\alpha)$,
\be
 \label{31}
 \psi' (\alpha) \equiv F'(\alpha)/ F(\alpha) \,, \qquad  G(\alpha)
 \equiv F^2(\alpha) - \third \Lambda e^{2\alpha}\,,
 \ee
 and use the following equations (the prime denotes differentiation
 w.r.t. $\alpha$):
\be
 \label{32}
 G' + 2G = \third \mu^2 A^2 \,, \qquad
 A'' + \psi' A' + \mu^2 e^{2\alpha} F^{-2} A = 0 \,.
 \ee
 Of course, instead of the first equation we can use the
 equivalent equation for  $F^2 \equiv \cF$ that directly
 follows from (\ref{29}):
 \be
 \label{33}
 \cF' + 2\cF = \4third \Lambda e^{2\alpha} +
 \third \mu^2 A^2 \,.
 \ee

 Together with the constraint  (\ref{30}), rewritten as
 \be
 \label{34}
 \Hag^a_1 =  -e^{2\alpha} \bigl[\cF (6 - e^{-2\alpha}A'^2 )
 -  2\Lambda e^{2\alpha} - \mu^2 A^2 \bigr] =0 \,,
 \ee
  equations (\ref{32}) form
 the complete system describing cosmology in the LC gauge.
 Note that the constraint (\ref{30}) is the integral of motion and thus
 it is sufficient to require that it vanishes just at one point, say,
 at $t=0$ or $\alpha = -\infty$.
 To derive possible asymptotic behaviour of the solutions
 for $|\alpha| \rightarrow \infty$ it is natural to expand $A$
 in powers of $e^{\alpha}$ and to self consistently use the general
 solution of the first equation,
 \be
 \label{35}
  G(\alpha) = \third \mu^2 e^{-2\alpha} \int d\alpha
  A^2 (\alpha) e^{2\alpha} \,,
  \ee
 with the relations for the expansion coefficients obtained from
 the second equation.

 In this way we can find, step by step, the asymptotic
 expansion. In the asymptotic region $\alpha  \rightarrow -\infty$
 we can then find the following possible asymptotic behaviour:
 \be
 \label{36}
 A = \sum_{n=0}^{\infty} A_n e^{\alpha n}\,,
 \qquad \cF = e^{-2\alpha } \biggl[C_{\infty} +
 \sum_{n=2}^{\infty} \cF_n e^{\alpha n} \biggr]\,,
 \qquad  \psi' = -1 + \sum_{n=2}^{\infty} n{\psi}_n e^{n \alpha} \,,
 \ee
 where $C_{\infty}$, $A_0$, $A_1$ are arbitrary constants\footnote{
 $A_1$ is defined by the constraint (\ref{30}):
 putting the first terms of the expansion (\ref{36}) into (\ref{30})
 or (\ref{34}) we get $A_1 = \sqrt 6$, which is sufficient for satisfying
 the constraint.}; $A_n$, $\cF_n$ for $n \geq 2$ are derived recursively
 from (\ref{32}), (\ref{35}), and ${\psi}_n$ from definition
 (\ref{31}). The first coefficients are:
 \be
 \label{37}
 A_1 = \pm \sqrt 6 , \,\,\,\, A_2 = 0 , \,\,\,\,
 A_3 = - \sixth {\frac{\mu^2 A_0^2}{A_1 C_{\infty}}}  \,
  , \,\,\,\,
 A_4 = {\frac{\mu^2 A_0 }{4 C_{\infty}}} \,;
 \ee
 \be
 \label{38}
 \cF_2 = \sixth \mu^2 A_0^2 \,, \quad
 \cF_3 = \frac{2}{9} \mu^2 A_0 A_1  \,;
 \qquad \psi_2 = {\frac{1}{2 C_{\infty}}} \cF_2
 \quad \psi_3 =  {\frac{1}{2 C_{\infty}}} \cF_3 \,.
 \ee
 Thus we find the differential equation for the metric function $f(t)$
 (`scale factor'):
 \be
 \label{39}
 \frac{d}{dt} \bigl(e^{\alpha}\bigr) \equiv \dot{f} =
 \sqrt{C_{\infty}} \, [ 1 + 2\psi_2 f^2 + 2\psi_3 f^3 + ... ]^{\half} \,,
  \ee
 and if we solve it we can find the vector field $A(t)$
 by using (\ref{36}), (\ref{37}). Neglecting the third term
 in the r.h.s.
 it is easy to solve this equation finding the dependence
 of $f$ on $t$:
 \bdm
 f(t) = {\frac {\sqrt{6 \, C_{\infty}}} {\mu A_0}}
 \sinh \biggl[ \biggl( \sixth \mu^2 A_0^2 \biggr)^{\half}
 (t-t_0) \biggr].
 \edm
 At first sight, the exponential growth of $f(t)$ suggests a possibility of an
 inflation character of this solution.
 However, this is only the first approximation and
 we should take into account higher order terms to get a more solid
 conclusion\footnote{
 An interesting exercise could be to keep four terms in the r.h.s. of
 (\ref{39}) and express the solution in terms of the elliptic functions.
 The behaviour of $f(t)$ in this approximation essentially
 depends on all the parameters.}.
  Moreover,  we see that the qualitative character of the
 solutions essentially depends on the physical parameters
 $A_0$, $\mu^2$ on which at the moment we have no reliable
 information\footnote{
 The dependence on $\Lambda$ only occurs in the omitted fourth-order
 terms.}.

 The discussed {\bf solution is not unique}.
 Using the above equations we can
 derive another one, for which both $A$ and $F$ are finite for
 $\alpha \rightarrow -\infty$. To get it we take
 \be
 \label{40}
 G(\alpha) = \third \mu^2 e^{-2\alpha} \int_{-\infty}^{\alpha} d\bar{\alpha}
  A^2 (\bar{\alpha}) e^{\bar{2\alpha}} \,,
 \ee
 and then apply the above procedure. Then, using the expansions
 \be
 \label{41}
 A = \sum_{n=0}^{\infty} A_n e^{2n\alpha} , \qquad
 F = \sum_{n=0}^{\infty} F_n e^{2n\alpha} , \qquad
 \cF = \sum_{n=0}^{\infty} \cF_n e^{2n\alpha} \,
 \ee
 we can find that
 \be
 \label{42}
 \cF_0 \equiv F_0^2 = \sixth \mu^2 A_0^2  \,, \qquad
 \cF_1 \equiv 2F_0 F_1
  = \biggl[ \third \Lambda + \sixth \mu^2 A_0 A_1 \biggr] \,,
 \ee
 \be
 \label{43}
 A_1 = -\mu^2 A_0 / 4F_0^2 = -3/2A_0 \,, \qquad
 A_2 = -A_1 (\mu^2 + 6 \cF_1 ) / 16 F_0^2 \,,
 \ee
 where now $A_0$ is the unique arbitrary constant
 (in the above solution we have one more constant
 $C_{\infty}$). Instead of Eq.(\ref{39}) we now have
 the equation:
 \be
 \label{44}
 F \equiv  \dot{\alpha}  = \bigl[ F_0^2  +
 2 F_1 F_0^{-1} e^{2\alpha(t)} + ...\bigr]^{\half} \,,
 \ee
 which can easily be solved in this approximation:
 \be
 \label{45}
 f \equiv e^{\alpha} = 2 e^{F_0 t} \bigl(1 -
 2 F_1 F_0^{-1} e^{2F_0 t}  \bigr)^{-1} \,.
 \ee
 The scale factor vanishes if $t \rightarrow -\infty$.
 The parameter $F_1 F_0^{-1}$ strongly depends on $A_0$, $\Lambda$,
 and $\mu^2$. It may be positive or negative, small or large:
 \be
 \label{46}
 F_1 F_0^{-1} = \biggl(\Lambda - \3quart \mu^2 \biggr)
 \biggl( \mu^2 A_0^2 \biggr)^{-1} \,.
 \ee
 We see that for the negative values of $F_1 F_0^{-1}$ the scale factor
 $f$ can grow with $t$ up to the maximum value $|F_0 F_1^{-1}|$
 and then decrease to zero. For positive $F_1 F_0^{-1}$ it blows
 up when the expression in the brackets vanishes, but then the approximation
 is certainly inapplicable. Thus, we must be very cautious
 in making definite conclusions basing on this simple result.
 The approximation (\ref{44}) can only be reasonable when
 $|F_1^{(2)}| f^2 \leq  F_0^2$. As $F_0$ and $F_1$
 strongly depend on $\Lambda$, on the absolutely unknown mass
  $\mu$ and on the arbitrary constant
 $A_0$, it is not possible (at the moment)
 to make conclusive statements on the general
 properties of this solution though it depends on less parameters
 than the first one. Note only that both solutions are compatible
 with existence of a period of fast growing scale factor.

 \subsubsection{A generalization}
 The expansions for both cosmological solutions can
 generally be written as follows. Let us use the  more
 convenient notation:
\be
 \label{ca1}
 \cF(\alpha) \equiv F^2 (\alpha) \equiv  C_{\infty} e^{-2\alpha} \,+\,
 \sum_{n=0}^{\infty} \cF_n e^{-n\alpha} \,,
 \qquad  \cA_n \equiv \sum_{l=0}^{n}
 A_l \, A_{n-l} \,, \qquad \bar{\Lambda} \equiv \Lambda /3 \,.
 \ee
 Then the general expression for $\cF_n$ can be written as
\be
 \label{ca2}
  \cF_n  \,=\,  \mu^2 \cA_n /\,[3(n+2)]  \,-\, k \, \delta_{n,0}
  \,+\,  \bar{\Lambda} \, \delta_{n,2} \,,
 \ee
 and Eq.(\ref{32}) for $A(\alpha) \equiv \sum_n A_n e^{n\alpha}$
 rewritten as
\be
 \label{ca3}
  2 \cF A'' \,+\, \cF' A' \,+\, 2\mu^2 e^{2\alpha} A  = 0   \,,
  \ee
 defines the recurrence relation for $n \geq 0$ ($A_n \equiv 0$
 for $n=-1, -2$):
 \be
 \label{ca4}
 2 C_{\infty} \, (n+1) \, (n+2) \, A_{n+2} \,+\, \sum_{m=1}^{n} m \, (m+n)
 \, \cF_{n-m} \, A_m \,+\, 2 \mu^2  A_{n-2} = 0 .
 \ee
 Taking $n=0$ we find that $C_{\infty} A_2 = 0$. The first solution
 discussed above is obtained when $A_2 = 0$, the second -- when
 $C_{\infty} = 0$ (in the second case
 $\cF_{2n+1} = A_{2n+1} = 0$ for all $n\geq 0$).
 In both cases $A_0$ is arbitrary
 and other $A_n$ are derived recursively from Eqs.(\ref{ca2}),
 (\ref{ca4}), taking into account that for the first solution
 $A_1$ is defined by the constraint (\ref{34}),
 i.e. $A_1 = \pm \surd 6$.

 Note that $C_{\infty} > 0$ but the signs of $A_0$ and of $A_1$
 are not fixed. For this reason, the structure
 of the series expansion of $A(\alpha)$ is rather complex.
 It follows that the behavior of $\alpha (t)$ with growing time
 may change its character (for example, depending on the parameters
 $C_{\infty}$, $A_0$, $\mu^2$, $\Lambda$, it may
 change the exponential growth to oscillating or even chaotic
  behavior).
 At the moment, the main unsolved problem is to derive the
 asymptotic behavior of $A_n$ and $\cF_n$ for
 $n \rightarrow \infty$. An educated guess is that the main terms
 are given by some powers of $n$ and thus the expansions in powers
 of $f = e^{\alpha}$ have a finite radius of convergence in the
 $f$-plane\footnote{This was proved for the analogous expansions near
 the horizons, see \cite{atfm}}.  If this is true, the radius of
 convergence and the precise position and nature of the
 corresponding singularity in the complex $f$-plane could give us
 a very important information on the solutions.

 The dimensionless functions $\cF(\alpha) /\mu^2$ and $A(\alpha)$
 of general solution depend
 on the dimensionless constants $\lambda \equiv \Lambda /\mu^2$,
 $\rho \equiv C_{\infty}/\mu^2$ and $A_0$. The dependence on $A_0$ is
 especially simple for the second solution. It is not difficult to prove
 (e.g., by induction) that
 \be
 \label{ca5}
 A_{2k} / A_0 = a_{2k} A_0^{-2k} \,, \qquad  \cF / \cF_0 =
 f_{2k} A_0^{-2k} \,,
 \qquad   \cF_0 \equiv \mu^2 A_0^2 /6  \,,
  \ee
 where $a_{2k}$,  $f_{2k}$ depend only on $\lambda$.
  We thus have the expansions
 \be
 \label{ca6}
 A(\alpha)/ A_0 = 1 + \sum_{k=1}^{\infty} a_{2k} \, e^{2k\bal} \,,
 \qquad \cF / \cF_0 = 1 + \sum_{k=1}^{\infty} f_{2k} \, e^{2k\bal} \,,
 \quad  \bal \equiv \alpha - \ln A_0 \,,
  \ee
 in which $a_{2k}$,  $f_{2k}$ can be derived by the recurrence relations
\be
 \label{ca7}
 f_{2n} = \third \lambda \delta_{1n} +
 \sum_{k=0}^n (n+1)^{-1} a_{2k} \, a_{2n -2k} \,,
 \qquad \sum_{k=1}^n k (n+k) f_{2n-2k} \, a_{2k} + 3a_{2n - 2}  = 0 \,.
 \ee
 It follows that the equation for $\alpha(t)$ can be rewritten as
 \be
 \label{ca8}
 \frac{d \bal}{d\tau} = \bigg[ 1 + \sum_{k=1}^{\infty}
  f_{2k} \, e^{2k \bal (\tau)} \biggr]^{\half} , \qquad
  \tau \equiv \mu A_0 / \sqrt 6\, t \,.
 \ee
 An interesting property of these expansions is the following: if the
 dimensionless parameter $A_0$ is large enough,
 $e^{\bal}$ may remain small even for large positive $\alpha$
 (thus the scale parameter $e^{\alpha}$ may be very large).
 This means that it may be possible to approximate the expansions by
 a small number of terms (in some domain of $t$ in which $e^{\alpha}$
 is very large). This property is of interest for discussing
 inflation models.
   The first several terms in the expansions are easy to derive
   ($\lambda \equiv \Lambda /\mu^2$):
\be
 \label{ca9}
 a_2 = -\frac{3}{2} \,, \qquad a_4 = {\frac{9}{8}}
 \biggl[\lambda - \quart \biggr] \,,
 \qquad  a_6 = -\frac{5}{4} \lambda \biggl[\lambda -
 \frac{21}{20} \biggr]\,;
 \ee
 \be
 \label{ca10}
 f_2 = 2 \biggl[\lambda - \3quart \biggr]  \,, \qquad
 f_4 = \3quart \biggl[\lambda + \3quart \biggr] \,,
 \qquad  f_6 = -\frac{5}{8} \biggl[\lambda - \frac{9}{20} \biggr]
 \biggl[\lambda + \3quart \biggr] \,.
 \ee

 Our simplified Lagrangian (\ref{26}) should be considered as
 a (mainly) mathematical model for a preliminary study of the  EW theory.
 To really discuss its cosmological applications
  we must first find an approximation valid for
 high values of $f$ and then return to the complete set of the
 equations of motion, to which the simplified model is only
 a rather crude approximation.
 We also should not forget that it is absolutely necessary to include
 into consideration `ordinary' matter before one can really
 discuss physical picture of the cosmological evolution.

 \section{Discussion}
 In this paper we briefly summarized the main ideas of the
 Einstein - Weyl model and presented its new interpretation,
 as well as some results obtained investigating its simplest solutions.
 We only considered the static spherically
 symmetric solutions and, in cosmology, only
 an artificially simplified homogeneous
 model. As we noted in \cite{ATF3}, even small deviations from the spherical
 symmetry may result in a qualitatively different theory. In particular, if
we consider axially symmetric configurations infinitesimally deviating
from the spherically symmetric ones, we will find additional scalar fields
in the vecton dilaton gravity, which may be very important in cosmological
 considerations and in analysing black holes. We did not touch these
 problems here. Moreover, even in the spherically symmetric case our
 study is incomplete.
 In the static case, we have only proven that there may exist
 two horizons and
 derived the solutions close to the horizons.
 In cosmology, we have studied only the asymptotic
 behavior of the solutions in the simplified model.

 As we mentioned above, we expect that
 the complete solutions should reveal some sort of
 chaotic behavior. To study these phenomena we must first
 carefully discuss
 the physical parameters of the theory. In the original formulation
 these are: the gravitational constant, the cosmological constant,
 and the vecton mass. In addition, the asymptotic boundary conditions
 introduce other parameters, the dependence on which is highly
 nontrivial. This does not allow us to make sound
  conclusions (or, even guesses) about the global behavior
 of the solutions derived in our essentially local approach. For example,
 if we try to glue together the left and the right asymptotic
 approximations, we will find that the gluing procedure is strongly dependent
 on the parameters that characterize the influence of the nonlinear terms in the
 equations, up to producing chaotic effects. This requires a
 very careful qualitative
 and numerical study of the equations.
 Of course, the most important task is taking into account the
 `ordinary' matter.

 Finally, we must admit that the vecton field is a rather unusual feature
 of the Einstein-Weyl model. I have found just a few papers
  in which a massive  vector field is introduced
  in the cosmological context (see, e.g.,
  \cite{Bertolami}-\cite{Armen}
  and references therein\footnote{
  In the second version of this paper I omitted two references in which
  the massive vector has only $A_0(t)$ component and thus
  $F_{ij} \equiv 0$.}).
 Thus the unified model of dark energy and vecton dark matter considered
 here looks as fresh and new as it was in 1923.
 In addition, I wish to stress that the original Einstein Lagrangian
 (\ref{3}) or (\ref{8}) is more interesting and exciting than the
 simplified theory (\ref{12}) (in particular, one may expect that in the
 original formulation there exists a relation among
 dimensional parameters that are arbitrary in the theory (\ref{12})).
 Unfortunately, the original theory is much more difficult
 to deal with and thus the prime
 goal must be the study of the simplified theory.
 In this paper I only give a sketch of how to begin such a study.

 {\bf Additional comments}

 {\bf 1.}~To better understand the brief exposure of the static
 solutions in Section~2.2 the reader is advised to consult our
 paper \cite{atfm}. There we proved that the expansions of
 the solutions represent analytic functions that analytically depend
 on all the parameters. This means that the two horizons defined by
 Eq.(\ref{25}) and corresponding to the same parameters
 $\tilde{A}_0$, $B_0$,  $\tilde{F}_0$, belong to the same static
 solution in spite of the fact that it is represented
 by two different expansions near the two horizons.
 Also, analyticity tells us that any solution can
 be analytically continued from one horizon to the other as well as
 to any regular point in the interval $0 < r < \infty$. The same
 remarks are also relevant to the cosmological solutions
 considered in Section~2.3. Recalling that by crossing the horizon
 we pass from the static to cosmological solutions one
 may even use the static coordinates for and alternative description
 of cosmologies (see, e.g., \cite{Erwin1}, \cite{Linde}).

 {\bf 2.}~Recently, several authors considered massive vector models
 of inflation (see, e.g., \cite{Mukhanov}-\cite{Germani} and
 references therein). As distinct from the EW model and from
 the first paper \cite{Ford} on the vector inflation,
 they introduce some non-minimal couplings
 of the vecton to gravity and/or many vector fields.
 In the Einstein approach additional vector fields
 (as well as the inflaton - type scalar fields) can
 possibly be produced in some higher-dimensional version.
 It is not clear whether non-minimal couplings can be produced
 but, possibly, similar effects could be imitated in the really
 non-linear Eddington-Einstein type theory. There exist attempts
 to use the non-linear Eddington-Einstein type actions for
 description of dark energy and dark matter (see, e.g.,
 \cite{Banados} and references therein).
  Note also that, possibly, the anisotropic effects of the
  vector field can be smeared
  out by inflation (see, e.g., \cite{Zeldovich} where the evolution
  of the universe with the primordial magnetic field was discussed).

  {\bf 3.}~In this paper, we did not discuss the possible interpretation
  of the vecton as a dark matter candidate. One reason for this is our
  inability to estimate its mass, even by order of magnitude.
  The other reason is that the vecton is interacting only with gravity
  and thus can be produced (in abundance) only in very high gravitational
  fields. This requires essentially quantum considerations that are beyond
  the scope of the present paper.

 \section{Appendix}
 The connection (\ref{5}) is a special case of the general expression for
 the connection in affine spaces.
The most {\bf general symmetric affine connection} has the form (see,
e.g., \cite{Norden}):
\be
 \label{a1}
\Gamma^m_{kl} = \half [s^{mn} (s_{nk,l} + s_{ln,k} - s_{kl,n}) +
 s^{mn} (s_{nkl} + s_{lnk} - s_{kln})] \,,
\ee
 where  $s_{kl}$ is an arbitrary symmetric tensor, $s^{mn}$ is the inverse
 matrix to $s_{kl}$, and $s_{kln}$ is an arbitrary tensor that is symmetric
 in $k$ and $l$. Both the Weyl and Einstein connections belong to the
  subclass for which $s_{kln}$ can be presented in the form:
\be
 \label{a2}
 s_{kln} = \alpha \, s_{kl}\, i_n + \beta ( s_{nk} \, i_l + s_{ln} \, i_k)
 \,.
\ee
 We may call it the {\bf Weyl-Einstein connection} (defining the Weyl-Einstein
 spaces).

 Inserting (\ref{a2}) into (\ref{a1}), we find:
\be
 \label{a3}
 \Gamma^m_{kl} = \half [s^{mn} (s_{nk,l} + s_{ln,k} - s_{kl,n}) +
 \alpha ( \delta^m_k \, i_l +  \delta^m_l \, i_k) -
   (\alpha - 2\beta) s_{kl}\, i^m ] \,.
\ee
 Now it is easy to find that the Einstein connection (\ref{5})
 corresponds to $\alpha = -\beta = \third$.
 The Weyl connection introduced in \cite{Weyl} corresponds
 to $\alpha = 1$, $\beta = 0$.
 I could not find a discussion of the geometry of spaces with
 the Einstein connection in accessible literature.
 The geometry of the Weyl spaces is considered in \cite{Ed}
 ($D=4$, Lorentzian signature), and in \cite{Norden} ($D=2,3,4$,
 Euclidean signature).

\bigskip
\bigskip
\bigskip

 {\bf Acknowledgment:}
 It is a great pleasure for the author to cordially thank for useful remarks
 V.~de~Alfaro, M.~Choptuik,  A.D.~Linde, V.A.~Rubakov, A.A.~Starobinsky,
 K.~Stelle, P.V.~Tretyakov, and E.~Witten.
 The author very much appreciates financial support from
 the Department of Theoretical Physics of the University of Turin and
 from INFN (Turin Section), where he began to work on this paper.
 This work was also supported in part by the Russian Foundation for Basic
 Research (Grant No. 06-01-00627-a).


\end{document}